\documentclass[a4paper,11pt]{article}
\usepackage[dvips]{graphicx}
\usepackage{amssymb}
\usepackage{amsmath}

\begin{document}

\title{Newton's Law Modifications due to a Sol Manifold Extra Dimensional Space}
\author{V.K.Oikonomou\thanks{
voiko@physics.auth.gr}\\
Technological Education Institute of Serres, \\ 62124 Serres, Greece\\
and\\
Dept. of Theoretical Physics Aristotle University of Thessaloniki,\\
Thessaloniki 541 24 Greece} \maketitle

\begin{abstract}
The corrections to the gravitational potential due to a Sol extra
dimensional compact manifold, denoted as $M_A^3$, are studied. The
total spacetime is $M^4\times M_A^3$. We compare the range of the
corrections to the range of the $T^3$ corrections. It is found
that for small values of the radius of the extra dimensions
($R<10^{-6}$) the the Sol manifold corrections are large compared
to the 3-torus corrections. Also, Sol manifolds corrections can be
larger, comparable or smaller compared to the 3-torus case, for
larger $R$.
\end{abstract}

\bigskip
\bigskip
\section*{Introduction}
In the last decade, the phenomenological implications of extra
dimensional models were extensively studied. Many of these studies
are concentrated on the modifications caused on the gravitational
potential due to extra dimensional manifolds
\cite{kehagias,leontaris,leontaris1}. Generally the modifications
are of Yukawa type (except in Randall-Sundrum warped extra
dimensional models), differing only in their strength and range
(for a very recent interesting application see \cite{leontaris1}).
We shall study the modification of the gravitational potential
caused by a Sol extra dimensional manifold. Sol manifolds are
compact and orientable three dimensional manifolds, denoted as
$M_A^3$. We shall present the Sol manifolds geometric structure.
Also we shall study the Laplace equation on such manifolds.
Finally we compare the range of the corrections to the
gravitational potential due to Sol manifolds, with the
corresponding range caused by a $T^3$ extra dimensional manifold.

\section*{Sol Manifold Geometry and Topology}
Sol manifolds are one of Thurston's 8 three dimensional geometries
\cite{Thurston} (geometric structures). The geometrization
conjecture is an approach to the geometry of three dimensional
manifolds through general topological arguments. One starts with
the fact that three dimensional manifolds are composed by
2-spheres or torii surfaces. According to the properties and the
details of the topological gluing map of the above with $R$ or
$S^1$, the resulting manifolds acquire locally homogeneous
metrics. Sol manifolds are described by one type of the 8
different homogeneous metrics.

Sol manifolds are obtained from $SL(2,Z)$ stiffennings of torus
bundles over the circle. A theorem \cite{tanimoto} states: given M
a total bundle space of a $T^2$ bundle over $S^1$ with gluing map
$\phi$ and let $A$ $\epsilon$ $GL(2,Z)$ represent the automorphism
of the fundamental group of the torus $T^2$ induced by the gluing
map $\phi$. Then the total bundle space admits a Sol geometric
structure if $A$ is hyperbolic, an $E^3$ structure if $A$ is
periodic and finally a Nil structure otherwise. $A$ is hyperbolic
and gives rise to an orientable manifold if $|TrA|>2$ and we shall
dwell on this choice. Thus Sol structure arises from the $SL(2,Z)$
stiffening of the mapping torus of a torus diffeomorphism $\phi$
\cite{Thurston}.

Two different hyperbolic gluings with $A_1\neq A_2$, give rise to
different geometric Sol structures. $A$ we will be represented
\begin{equation}\label{ASTRU}
A=\left(%
\begin{array}{cc}
  0 & 1 \\
  -1 & n \\
\end{array}%
\right)
\end{equation}
so the different Sol structures are classified by the integer $n$,
with $|n|=|TrA|>2$. Also only the positive eigenvalues of $A$.

Let us analyze more the Sol manifold structure. Consider the
manifold $T^2\times \mathbb{R}$ described by the periodic
coordinates $(x,y)$ for the torus, defined modulo $R$ (the radius
of the compact dimensions) and $z$ $\epsilon$ $(-\infty,\infty)$
be a coordinate of $\mathbb{R}$. The combined action of the torus
mapping through the hyperbolic gluing map $A$ diffeomorphism
(which we denote as $\bar{\Gamma}$) is :
\begin{equation}\label{mapptorus}
\bar{\Gamma}:\left(
\begin{array}{c}
  x \\
  y \\
  z \\
\end{array}
\right)
\rightarrow \left(%
\begin{array}{c}
  a_{11}x+a_{12}y \\
  a_{21}x+a_{22}y \\
  z+2\pi R \\
\end{array}%
\right)
\end{equation}
Let the matrix $A$, be of the form
\begin{equation}\label{generalm}
    A=\left(%
\begin{array}{cc}
  a_{11} & a_{12} \\
  a_{21} & a_{22} \\
\end{array}%
\right)
\end{equation}
but in practise we shall use the form of relation (\ref{ASTRU}).
The Sol manifold, which is denoted as $M_A^3$ is the quotient of
$T^2\times \mathbb{R}$ by the action of $\bar{\Gamma}$, that is
$M_A^3\equiv T^2\times \mathbb{R}/\bar{\Gamma}$. It is a total
torus bundle over $S^1$, with $T^2$ the fiber, $S^1$ the base
space and $A$ the hyperbolic gluing map of the torus fibers. We
consider Sol manifolds for which the eigenvalues $\lambda$ of $A$
are positive. For the case of (\ref{ASTRU}) the characteristic
polynomial of the matrix $A$ reads:
\begin{equation}\label{charpoly}
\lambda^2-TrA\lambda +1=0
\end{equation}
or equivalently:
\begin{equation}\label{charpol1}
\lambda^2-n\lambda +1=0
\end{equation}
The solutions to equation (\ref{charpol1}) are $\lambda$ and
$\lambda^{-1}$ with $\lambda+\lambda^{-1}=n=TrA$. Also the
discriminant of $A$ is:
\begin{equation}\label{discriminant}
D=(a_{11}+a_{22})^2-4
\end{equation}
which in our case reads:
\begin{equation}\label{discri1}
D=(\lambda -\lambda^{-1})^2
\end{equation}
We use another coordinate system $(u,v,z)$ on the Sol manifold.
The $(u,v)$ are linear coordinates of the torus fibres related to
an eigenbasis of the hyperbolic map $A$. These coordinates
correspond to a rotated torus lattice. The action of
$\bar{\Gamma}$ in the new coordinate system reads:
\begin{equation}\label{newcoor}
\bar{\Gamma}:\left(
\begin{array}{c}
  u \\
  v \\
  z \\
\end{array}
\right)
\rightarrow \left(%
\begin{array}{c}
  \lambda u \\
  \lambda^{-1}v \\
  z+2\pi R \\
\end{array}%
\right)
\end{equation}
The original lattice was orthogonal while the new torus lattice is
not. If $e_u$ and $e_v$ are the basis of the lattice after the
action of $\bar{\Gamma}$ identification, then
$(e_u,e_v)=|e_u||e_v|\cos \theta$. Also the fiber coordinates
$T^2$ are not periodic anymore. So in order two pairs $(u_1,v_1)$
and $(u_2,v_2)$ define the same point on the torus lattice, the
following must hold:
\begin{equation}\label{periodi}
(u_1-u_2,v_1-v_2)=ke_u+me_v
\end{equation}
with $k,m$ integers and $e_u$, $e_v$ defined previously.

\subsubsection*{Riemannian Sol group invariant metric on Sol
manifolds}

The Riemannian metric on Sol manifolds comes from the metric on
the universal covering of $M_A^3$. The group invariant metric on
the universal covering is a Sol group invariant metric. With the
invariant metric on the universal covering of $M_A^3$ we can find
the following class of metrics of Sol manifolds:
\begin{equation}\label{metricsol}
\mathrm{d}s^2=Ee^{2z\ln\lambda}\mathrm{d}u^2+2F\mathrm{d}u\mathrm{d}v+Ge^{-2z\ln\lambda}\mathrm{d}v^2+\mathrm{d}z^2
\end{equation}
We shall use the metric (\ref{metricsol}) in order to compute the
Laplace-Beltrami operator on $M_A^3$.

\noindent Sol manifolds are hyperbolic manifolds with negative
curvature. The hyperbolicity is a very interesting feature of Sol
manifolds, regarding they are torus fibrations. Due to their
hyperbolicity, the distribution of eigenvalues of the Laplacian is
not Poisson. This might have effects on the gravitons mass
splitting.
\section*{Newton's law and extra dimensions}

Our purpose is to examine the corrections to the gravitational
potential caused by an extra dimensional Sol manifold. The
spacetime manifold is of the form $M^4\times M_A^3$. Let us review
the general technique to obtain these corrections. The
presentation is based on reference \cite{kehagias,oikonomou}.

Consider a spacetime of the form $M^4\times M^n$, with $M^n$ an
n-dimensional compact manifold and $M^4$ the four dimensional
Minkowski spacetime and also a complete set of orthogonal harmonic
functions on $M^4$, $\Psi_m$, satisfying the orthogonality
condition:
\begin{equation}\label{orthogo}
\int_{M^n}\Psi_n(x)\Psi_m^*(x)=\delta_{n,m}
\end{equation}
and the completeness relation:
\begin{equation}\label{complet}
\sum_m\Psi_m(x)\Psi_m^*(x')=\delta^{(n)}(x-x')
\end{equation}
The functions $\Psi_m$ are eigenfunctions of the $n$-dimensional
Laplace-Beltrami operator $\Delta_n$ of the manifold $M^n$, with
eigenvalues $\mu_m^2$:
\begin{equation}\label{eigenvalueeq}
-\Delta_n\Psi_m=\mu_m^2\Psi_m
\end{equation}
The gravitational potential $V_{n+4}$ satisfies the Poisson
equation in $n+3$ spatial dimensions, when the Newtonian limit is
taken:
\begin{equation}\label{lap}
\Delta_{n+3}V_{n+4}=(n+1)\Omega_{n+2}G_{n+4}M\delta^{(n+3)}(x)
\end{equation}
with $M$, the mass of the system, $G_{n+4}$ the Newton constant in
$n+4$ dimensions and
\begin{equation}\label{omega}
\Omega_{n+2}=\frac{2\pi^{\frac{n+3}{2}}}{\Gamma(\frac{n+3}{2})}
\end{equation}
Equation (\ref{lap}) corresponds to the case of large compact
radius limit and has the solution:
\begin{equation}\label{neweq}
V_{n+4}=-\frac{G_{n+4}M}{r_n^{n+1}}
\end{equation}
When the compact dimensions have small lengths, we find the
harmonic expansion of $V_{n+4}$ in terms of the eigenfunctions of
the product space $M^4\times M^n$, which reads:
\begin{equation}\label{harmonicexpansion}
V_{n+4}=\sum_m\Phi_m(r)\Psi_m(x)
\end{equation}
with $r$ denoting the coordinates of $M^4$ and $x$ denoting the
coordinates of $M^n$. Consequently, the $\Phi_m$ obey:
\begin{equation}\label{poiss3}
\Delta_3\Phi_m-\mu_m^2\Phi_m=(n+1)\Omega_{n+2}\Psi_m^*(0)G_{n+4}M\delta^{3}(x)
\end{equation}
with solution:
\begin{equation}\label{gravpot12}
\Phi_m(r)=-\frac{\Omega_nG_{n+4}M\Psi_m^*(0)}{2}\frac{e^{-|\mu_m|r}}{r}
\end{equation}
The gravitational potential is written as:
\begin{equation}\label{fingravpot}
V_{n+4}=-\frac{\Omega_nG_{n+4}M}{2r}\sum_m\Psi_m^*(0)\Psi_m(x)e^{-|\mu_m|r}
\end{equation}
Since all point particles in the four dimensional spacetime have
no dependence on the internal compact space $M^n$, we can take
$x=0$ in (\ref{fingravpot}) to obtain the four dimensional
gravitational potential:
\begin{equation}\label{fingravpot1}
V_{4}=-\frac{G_{4}M}{r}\sum_m\Psi_m^*(0)\Psi_m(0)e^{-|\mu_m|r}
\end{equation}
which is valid for large values of $r$, compared to the lengths of
the compact dimensions.

In the above general result of relation (\ref{fingravpot1}) we
shall apply the eigenfunctions and eigenvalues of Sol manifolds.

\section*{Sol Manifold modification of Newton's Law}
In order to compute the corrections to the gravitational potential
we must solve equation (\ref{eigenvalueeq}) for the case of Sol
manifold $M_A^3$. A much more elaborate analysis of this section
can be found in \cite{bolsinov,oikonomou}. Using the $(u,v,z)$
coordinates we introduced previously, the Laplace-Beltrami
operator for the manifold $M^3_A$ is:
\begin{equation}\label{Laplaceonsol}
\Delta=Ee^{2z\ln\lambda}\frac{\partial^2}{\partial
u^2}+2F\frac{\partial^2}{\partial u\partial
v}+Ge^{-2z\ln\lambda}\frac{\partial^2}{\partial
v^2}+\frac{\partial^2}{\partial z^2}
\end{equation}
which stems from the Riemannian metric (\ref{metricsol}). As usual
$E=|e_u|^2$, $F=|e_v|^2$ and $G=|(e_u,e_v)|$, where $e_u$ and
$e_v$ the basis of the $T^2$ lattice. Thus we must solve
\begin{equation}\label{laplacegener}
-\Delta \psi=E\psi
\end{equation}
A function $\Psi=e^{2\pi i(\gamma,{\,}w)}f(z)$ satisfies equation
(\ref{laplacegener}) if and only if $f(z)$ satisfies the modified
Mathieu equation \cite{oikonomou}:
\begin{equation}\label{modmahtieu}
\Big{(}-\frac{d^2}{dz^2}+|\nu(\gamma)|\cosh2\mu(z+\alpha(\gamma))\Big{)}f(z)=\Lambda
f(z)
\end{equation}
with $\mu=\ln \lambda$, $\nu(\gamma)=8\pi^2CQ_{A^*}(\gamma)$ and
$\alpha(\gamma)=\frac{\ln\big{(}\sqrt{\frac{E}{G}}\frac{|(\gamma,e_u)|}{|(\gamma,e_v)|}\big{)}}{2\ln\lambda}$.
Also
\begin{equation}\label{c}
C=\frac{1}{\sqrt{D}\mathrm{Vol}(M_A^3)\sin\theta}
\end{equation}
The volume of the Sol manifold, ${Vol}(M_A^3)$, is equal to the
volume of the total bundle space $T^2\times S^1$. $D$ is the
discriminant of the matrix $A$ defined in (\ref{discri1}) and
$Q_{A^*}(\gamma)$ is the quadratic form (see \cite{bolsinov})
corresponding to $A^*$ acting to the dual lattice of $T^2$, with
$Q_{A^*}(\gamma)=(\gamma,e_u)(\gamma,e_v)(\lambda-\lambda^{-1})$.
The eigenvalues $E$ and $\Lambda$ are related as follows:
\begin{equation}\label{eigenf}
E=\Lambda+\nu(\gamma)\cos\theta
\end{equation}
It is proved in reference \cite{bolsinov} that the functions
$\Psi_{\gamma}=e^{2\pi i(\gamma,{\,}w)}f_{\gamma}(z)$ form a
complete orthogonal basis on the Sol manifold. The spectrum of the
Laplace-Beltrami operator consists of two parts:

\begin{itemize}
\item The trivial part, with eigenvalues
$E_k=\frac{4\pi^2k^2}{R^2}$

\item The non-trivial part with eigenvalues
$E_{k,[\gamma]}=\Lambda_k(\nu[\gamma])+\nu([\gamma])\cos\theta$
and eigenfunctions the solutions of (\ref{modmahtieu}).
\end{itemize}

Thus we must compute the first eigenvalues and eigenfunctions. We
shall use the most interesting case which is when
$Vol(M_A^3)\sin\theta$ is large (or equivalently $\nu\rightarrow
0$). This term contains the compactification radius of the extra
dimensions and the deformation of the lattice in terms of
$\theta$. When $Vol(M_A^3)\sin\theta$ becomes large, $\nu(\gamma)$
becomes small \cite{bolsinov}. The eigenvalues of the
Laplace-Beltrami operator are $E_k=\Lambda_k$, with $\Lambda_k$
the eigenvalues of the modified Mathieu operator,
\begin{equation}\label{modopoe}
M=-\frac{d^2}{dz^2}+|\nu(\gamma)|\cosh2\mu z
\end{equation}
Let the order of the eigenvalues be $E_0=0\leq E_1\leq
E_2\leq...$. We shall use the first two, since the corrections to
the gravitational potential fall exponentially at the eigenvalues
grow larger. Thus the first two are $E_0=0$ and $E_1$, which
asymptotically read:
\begin{equation}\label{firsteigenvalue}
E_1\sim \frac{(\ln\lambda)^2\pi^2}{(\ln C)^2}
\end{equation}
with $C$:
\begin{equation}\label{c1}
C=\frac{1}{\sqrt{D}\mathrm{Vol}(M_A^3)\sin\theta}
\end{equation}
The eigenfunctions corresponding to the Mathieu operator
(\ref{modopoe}) are,
\begin{equation}\label{momathfunc}
f_{m}(z)=\sum_{k=0}^{\infty}A_{2k}^{2m}\cosh[2kz]
\end{equation}
with $'m'$ counting the eigenvalues, $m=0$ corresponds to $E_0$
e.t.c.

We substitute $E_0$ and $E_1$ in relation (\ref{fingravpot1}) with
$|\mu_0|=\sqrt{E_0}$ and $|\mu_1|=\sqrt{E_1}$. Also we substitute
the eigenfunctions of the Sol manifold, thus:
\begin{equation}\label{eigenzero}
 \Psi_m(0)=\sum_{k=0}^{\infty}A_{2k}^{2m}
\end{equation}
The asymptotic behavior of the coefficients $A_{2k}^{2m}$ for
$\nu\rightarrow 0$ is really simple \cite{mclachan}. The terms of
the form $A_m^{(m)}$ tend to $1$, while terms of the form
$A_m^{(k)}$, with $m\neq k$ tend to zero and the gravitational
potential of relation (\ref{fingravpot1}) becomes:
\begin{equation}\label{fingravpot13}
V_{4}=-\frac{G_{4}M}{r}\sum_m\sum_{k=0}^{\infty}A_{2k}^{2m}A_{2k}^{2m}e^{-|\mu_m|r}
\end{equation}
For the first two eigenvalues we have
\begin{equation}\label{lastexpr}
V_{4}=-\frac{G_{4}M}{r}\Big{(}A_{0}^{0}A_{0}^{0}e^{-|E_0|r}+A_{2}^{2}A_{2}^{2}e^{-|E_1|r}\Big{)}
\end{equation}
or (using the asymptotic behavior for the coefficients
$A_m^{(m)}$):
\begin{equation}\label{lastexpr1}
V_{4}=-\frac{G_{4}M}{r}\Big{(}e^{-\sqrt{E_0}r}+e^{-\sqrt{E_1}r}\Big{)}
\end{equation}
Since $E_0=0$ and using relations (\ref{firsteigenvalue}) and
(\ref{c1}) we obtain finally:
\begin{equation}\label{lastexpr2}
V_{4}\sim
-\frac{G_{4}M}{r}\Big{(}1+e^{-\big{|}\frac{(\ln\lambda)\pi}{(\ln
C)}\big{|}r}\Big{)}
\end{equation}
It is clear that the Sol manifold correction to the Newton's law
gravitational potential depends on the compactification radii of
the extra dimensions $R$, the angle $\theta$ of the vectors $e_u$
and $e_v$ and on the eigenvalues of the hyperbolic gluing map $A$.
In the next section we study the parameter space of the
corrections found and we compare it with the $T^3$ manifold
corrections.

\section*{Analysis of the parameter space and comparison with the
$T^3$ corrections}

\begin{figure}[h]
\begin{center}
\includegraphics[scale=.7]{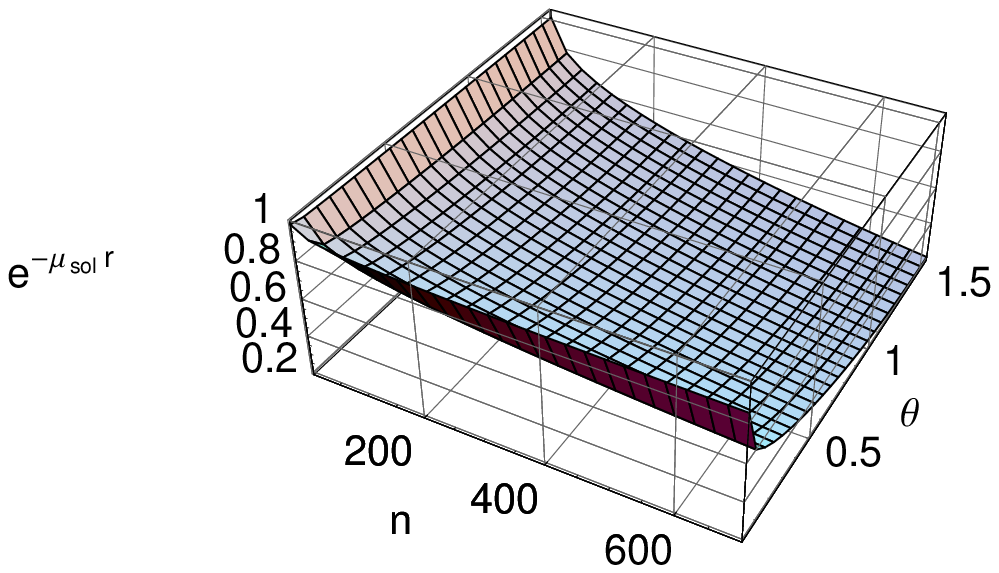}
\end{center}
\caption{Dependence of the range $e^{-\mu_mr}$ of Sol corrections
on $\theta$ and $n$ with $r=0.1{\,}mm$, $R=0.05{\,}mm$}
\label{3d1}
\end{figure}

\begin{figure}[h]
\begin{center}
\includegraphics[scale=.7]{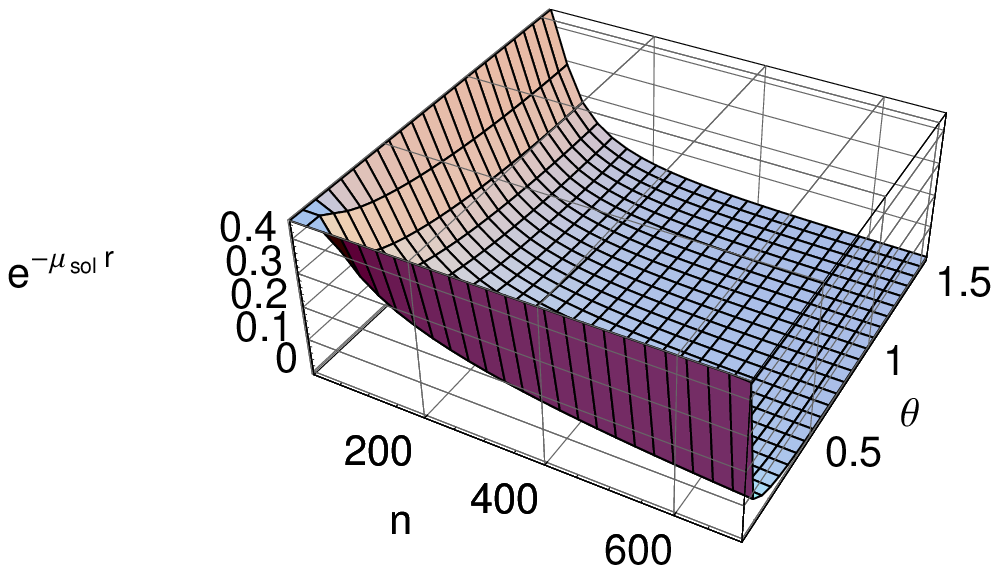}
\end{center}
\caption{Dependence of the range $e^{-\mu_mr}$ of Sol corrections
on $\theta$ and $n$ with $r=0.1{\,}mm$, $R=0.1{\,}mm$}\label{3d2}
\end{figure}

\begin{figure}[h]
\begin{center}
\includegraphics[scale=.8]{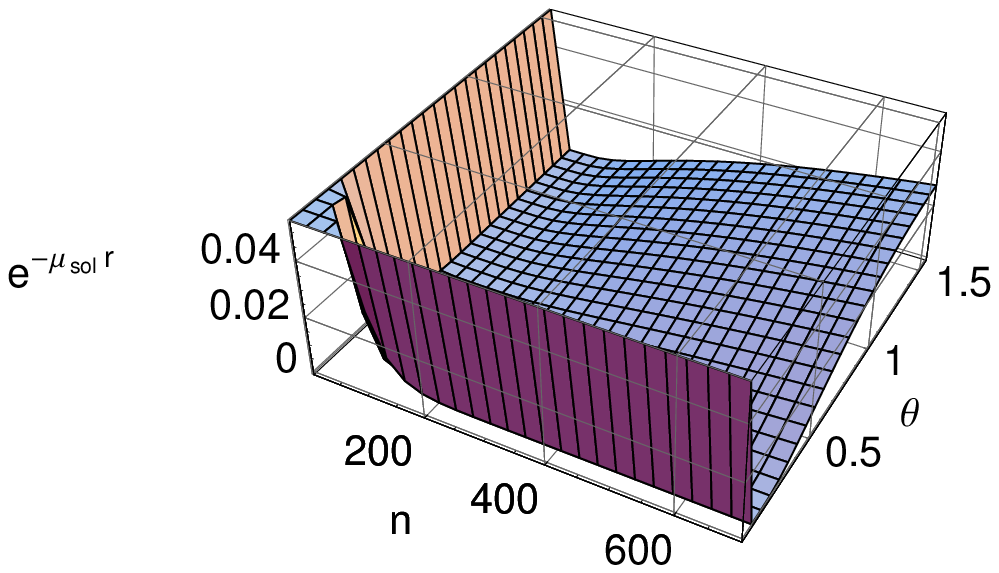}
\end{center}
\caption{Dependence of the range $e^{-\mu_mr}$ of Sol corrections
on $\theta$ and $n$ with $r=0.1{\,}mm$, $R=0.2{\,}mm$}\label{0201}
\end{figure}

We examine first the dependence of the Sol correction range
$e^{-\mu_mr}$ on the parameters $\theta$ and $n$. In Figure
(\ref{3d1}) we plot the dependence for the compactification radius
value $R=0.05{\,}mm$ (smaller than the current experimental bound
$r=0.2mm$) and for $r=0.1{\,}mm$, while in Figure (\ref{3d2}) and
(\ref{0201}) the values of $R$ are $0.1{\,}mm$ and $0.2{\,}mm$
respectively. The Sol structure gives very large corrections to
gravity if the compactification radius is very small (of order
$R\sim 10^{-6}{\,}$ and smaller). Also we shall check out the
behavior of the range of the corrections around $R\sim 10^{-8}m$
which is the expected scale that three extra dimensional spaces
should have. According to Figures (\ref{3d1}), (\ref{3d2}), and
(\ref{0201}), we can see that the last two are similar. Thus for
large values of $n$ ($>200$) and for $\theta>0.3$, the corrections
are very small. In the other two graphs when $n>300$ and for small
values of $\theta$, the corrections are small. We shall compare
the range of Sol manifolds with the range of the $T^3$ torus
corrections.

\begin{figure}[h]
\begin{center}
\includegraphics[scale=.7]{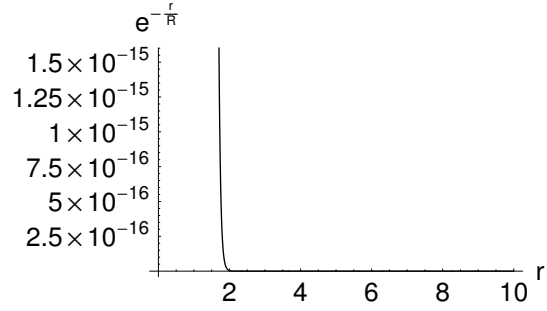}
\end{center}
\caption{$r$-dependence of the range of $T^3$ corrections for
$R=0.05{\,}mm$}\label{t3}
\end{figure}

\begin{figure}[h]
\begin{center}
\includegraphics[scale=.7]{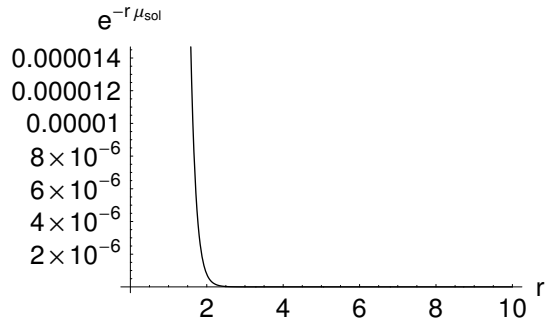}
\end{center}
\caption{$r$-dependence of the range $e^{-\mu_mr}$ of Sol
corrections with $R=0.05{\,}mm$, $n=250$ and
$\theta=\frac{\pi}{3}$}\label{005}
\end{figure}

 In Figure (\ref{t3}) we plot the range of the $T^3$ corrections as a function of $r$ with $R=0.05mm$ and
$\theta=\pi/3$ and in Figure (\ref{005}) the corresponding
dependence for the Sol manifold case. As it can be seen from
Figure (\ref{3d1}), the term $e^{-\mu_mr}$ is very big compared to
$e^{-r/R}$, for small $n$ ($\sim 5$). As $n$ grows the term
becomes smaller and smaller. Figure (\ref{005}) corresponds to
$n=250$. As $n$ grows, the range of Sol-corrections becomes
comparable and after a value of $n$, smaller compared to the $T^3$
corrections.

\begin{figure}[h]
\begin{center}
\includegraphics[scale=.8]{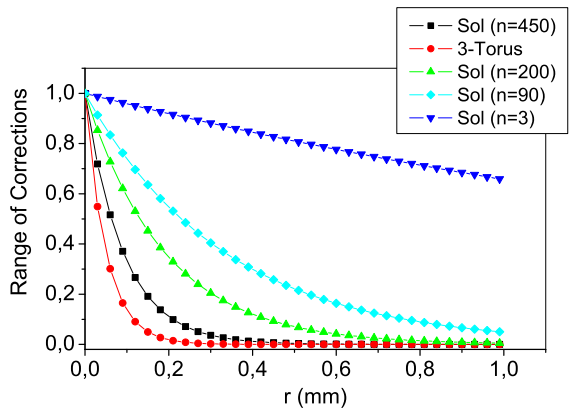}
\end{center}
\caption{Comparison of the $r$-dependence of the range of $T^3$
and Sol corrections for $R=0.2{\,}${\,}mm for various $n$, and
$\theta=\pi/3$} \label{comparissonnew}
\end{figure}

The case with $R=0.2${\,}mm is very interesting. According to
Figure (\ref{comparissonnew}) the range of Sol corrections can
vary significantly depending on the value that $n$ takes. Compared
to the corresponding 3-Torus range, Sol corrections can be much
larger (for small $n$) and comparable (for $n>450$) but never
smaller even for very large $n$.

Also Sol corrections can exist for compactification radii and at
distances for which the corresponding $T^3$ torus range values are
very large (and consequently not experimentally preferable).

\section*{Conclusions}
We studied the corrections to the gravitational potential caused
by a Sol manifold extra dimensional compact space. After
investigating the parameter space we found that the range of Sol
manifolds corrections can be similar to the $T^3$ torus results
and can be very different compared to the $T^3$ results depending
on the values of the parameters.

We left unanswered two issues, the graviton production and how do
we distinguish Sol manifolds corrections from other corrections
(the last due to the rich parameter space of Sol manifolds).

\section*{Acknowledgements}
V.O would like to thank Professor G. K. Leontaris for the
hospitality at the University of Ioannina.

\newpage

\newpage

\end{document}